\documentstyle[preprint,aps]{revtex}
\begin{document}
\title {Factors Responsible for the Stability and the Existence of a Clean 
Energy Gap of a Silicon Nanocluster}

\author{Lei Liu, C. S. Jayanthi, and Shi-Yu Wu}
\address{ Department of Physics, University of Louisville, 
Louisville, KY 40292}

\date{\today}
\maketitle

\begin{abstract}

We present a critical theoretical study of electronic properties of 
silicon nanoclusters, in particular
the roles played by symmetry, relaxation, and hydrogen passivation on the 
the stability, the gap states and the energy gap of the system using 
the order-N [O(N)] non-orthogonal 
tight-binding molecular dynamics and the local analysis of electronic structure.
We find that for an unrelaxed cluster with its
atoms occupying the regular tetrahedral network, the presence of 
undistorted local bonding configuration
is sufficient for the appearance of a small clean energy gap. 
However, the energy gap of the
unrelaxed cluster does not start at the highest occupied 
molecular orbital (HOMO). In fact, between
the HOMO and the lower-edge of the energy gap, localized dangling bond 
states are found. 
With hydrogen passivation, the localized dangling bond states 
are eliminated, resulting in a
wider and clean energy gap. Relaxation of these hydrogen 
passivated clusters does not alter 
either the structure or the energy gap appreciably. 
However, if the silicon clusters are allowed to
relax first, the majority of the dangling bonds are eliminated 
but additional defect states due to bond 
distortion appear, making the energy gap dirty. Hydrogen 
passivation of these relaxed clusters will
further eliminate most of the remnant dangling bonds but no 
appreciable effect on the defect states associated
with bond distortions, thus still resulting in a dirty gap. 
For the hydrogen-passivated $Si_N$ nanoclusters 
with no bond distortion and no overall symmetry, we have 
studied the variation of the energy gap as a function 
of size of the cluster for $N$ in the range of $80<N<6000$. 
The dependence of 
the energy gap on the size shows a similar behavior as that for 
silicon nanoclusters with no bond distortion 
but possessing overall symmetry.

\end{abstract}

\newpage
\section{Introduction}
\label{intro}

The discovery of photoluminescence (PL) in the 
visible regime \cite{LTC90} from porous silicon has 
prompted many experimental \cite{EXP91,EXP92,EXP92a,EXP93} 
and theoretical investigations 
\cite{SYR92,JPP92,BD93,MH93,MH94,TU94,AZ94}, both because of
the implications of this phenomenon in the fabrication of 
silicon-based optoelectronic devices and the
interest in understanding this phenomenon.  The intensive 
research following
this discovery seemed to suggest that strong optical 
transitions can be expected only from those
Si-based systems which have an effective reduced 
dimensionality. Structural studies of porous
Si showed that porous Si is composed of Si nanostructures 
in the forms of columns and clusters
\cite{EXP92,EXP92a,KL92}.
Hence, the ensuing research efforts have placed emphasis 
on the electronic and optical properties
of silicon nanoclusters. Two mechanisms have been proposed 
to explain the observation of 
enhanced PL. The first suggests that the strong 
PL in the visible 
regime is due to the enhancement of the momentum matrix 
elements associated with the confinement 
of the electronic wavefunctions \cite{GDS92,CD93,MSH94,AJR92} 
of silicon nanoparticles. The second
suggests that the surface chemical composition of silicon 
nanostructures may also play a key role
in the enhanced PL \cite{MSB92,FK92,SMP93}. In both 
mechanisms proposed, 
the existence of a clean energy gap plays a pivotal role in 
determining the efficiency of the luminescence. 
Furthermore, since the PL for the bulk crystalline silicon is 
in the near infrared regime while 
it is in the visible regime for silicon nanoparticles, a great 
deal of effort has been devoted to the study
of size 
dependence of the gap of Si 
nanoparticles \cite{SYR92,JPP92,BD93,MH93,MH94,TU94,AZ94}.

In the present work, we focus on issues that are relevant to 
the fabrication of device-quality
silicon nanostructures. Experimentally, it is difficult to 
control the number of atoms in a silicon nanocluster. 
In most cases, the experimentally prepared samples of 
silicon nano-clusters may not possess the full 
tetrahedral symmetry of the bulk crystal. However, the 
previous theoretical studies, particularly those
pertaining to the size-dependence of the energy gap, 
focused on highly symmetric configuration
where the surface dangling bonds are completely
passivated by hydrogen atoms so that the system possesses 
the regular tetrahedral bonding configuration. 
In this work, we will compare the results for nanoclusters 
with overall
symmetry to those with no overall symmetry. We will consider 
both hydrogen-passivated clusters and those
with no hydrogen passivation. Finally, we will consider both 
unrelaxed and relaxed nanoclusters. More specifically,
we will address the following issues: i) How to obtain a 
clean energy gap in the case of silicon 
nanoclusters? (ii) What is the nature of the gap states? 
(iii) What is the role of relaxation and hydrogen 
passivation on the energy gap and gap states? (iv) Is there 
an intrinsic relationship between the energy gap and 
the local bonding configuration? and (v) How is a 
nanocluster with full symmetry different from a cluster
with no overall symmetry with regard to the existence of 
a clean gap?

In this work we have used a combination of theoretical studies 
to investigate the structure, the stability,
the energy gap, and the electronic density of states of 
the nanocluster. We use the non-orthogonal 
tight-binding (NOTB) molecular dynamics \cite{MM97} to 
determine the structure of silicon nanoclusters. 
For sizes of the cluster with $n > 500$, we have used 
the order-N [O(N)] technique, as
developed in Ref. \cite{CSJ98}, to determine the equilibrium 
structure. Similarly, to circumvent
the difficulties associated with an accurate calculation of 
the energy gap of a large cluster, 
we have devised a method, as described in the Appendix of 
this paper, that directly computes the eigenvalues
corresponding to the highest occupied molecular orbital (HOMO) 
and the lowest unoccupied molecular orbital 
(LUMO) without having to obtain the entire eigenvalue 
spectrum. Finally, the electronic density of states are 
calculated using the Real-Space Green's function (RSGF) 
technique as developed in Ref. \cite{SYW95}. 

This paper is organized as follows. In sec. II, we 
consider Si nanoclusters with no overall
tetrahedral symmetry (e.g. $Si_{200}$, $Si_{800}$, 
and $Si_{2000}$) and 
illustrate how the relaxation of such systems, 
while leading to the
stability of the system, precludes the formation of 
a clean energy gap. In sec. III, we consider 
a silicon nanocluster with overall tetrahedral symmetry 
(e.g. $Si_{83})$ and investigate in
detail the role of hydrogen passivation on the stability, 
the energy gap, and gap states of such a cluster.
In sec. IV, we consider the hydrogen passivation of 
clusters with no overall tetrahedral
symmetry and compare our results with the case discussed 
in sec. III.  In sec. V, the variation of the energy
gap as a function of the size of the hydrogen-passivated 
nanocluster with no overall symmetry is given. Finally, 
in the Appendix, we 
present the formalism used in this paper for the 
calculation of HOMO and LUMO energies of a large cluster.

\section {Interplay Between Relaxation, Stability,and Energy Gap}

A nano-cluster, left to itself, will search for 
a stable configuration
by relaxation and "surface" reconstruction. The 
process is triggered by the 
tendency of a cluster to minimize its total energy 
with the elimination of the
dangling bonds associated with the exterior atoms. 
But it often comes at a 
price as it causes severe distortion of the bonds, 
in particular
of the exterior atoms \cite{EK97}. As a result, defect
states associated with bond distortions are expected 
to appear. Furthermore, 
relaxation may not completely eliminate the dangling 
bonds on the surface. Hence there will
also be defect states associated with remnant dangling 
bonds. Both these 
defect states may fall in the gap region, making the 
gap less clean or
even "dirty".

To verify the validity of this picture, we have 
performed the full-geometry
optimization of Si nanoclusters of different 
sizes and then
determine their EDOS. In our calculation, we
have deliberately chosen to study $Si_{200}$, 
$Si_{800}$, and $Si_{2000}$. 
The number of atoms in these clusters is such 
that they form incomplete
shells when the atoms are made to occupy the 
regular tetrahedral network
sites. Therefore, such clusters will not possess 
the overall tetrahedral 
symmetry. The NOTB Hamiltonian used in this 
work \cite{MM97} has been extensively
used in the literature and is known to yield 
reliable structures
for Si clusters \cite{DRA99} and Si(100) 
surface \cite{SL99}.  
When the number of atoms ($N$) in the 
cluster exceeds 500, we have used
the order-N technique as developed in 
Ref. \cite {CSJ98} to overcome the $N^3$ 
bottleneck in the computation of total 
energy and atomic forces. 

In our simulations, a stable configuration 
for a cluster of a 
given size is considered to have been reached 
if the magnitude of the force
acting on each atom is of the 
order $10^{-2} eV/\AA$ and the total
energy of the configuration is a minimum. 
Fig.1 shows the 
optimized structure for Si$_{2000}$. It can 
be seen that the interior of
the cluster has the bulk-like tetrahedral 
structure. On the 
"surface" of this cluster,  features resembling 
the adatoms in the Si(111)7x7 
reconstruction,\cite{KT85}
and dimer rows associated with Si(100)\cite{SL99} 
have emerged. The general features of the 
other clusters studied are similar to 
those characterizing the Si$_{2000}$
cluster, namely, the bonding configurations 
of the interior atoms are
bulk-like and the surface features exhibit 
the characteristics of the most
stable surface reconstruction of Si \cite{CSJ98}. 
However, the bonding configurations
of exterior atoms show more severe distortions 
from the regular tetrahedral 
network as the size of the cluster decreases.

A survey of the exterior atoms in these 
clusters (see Fig.1)
shows that most of them are 3-fold bonded. 
Thus, most of the dangling 
bonds associated with the unrelaxed exterior 
atoms have been eliminated
by the surface reconstruction, specifically 
the tendency of the Si atoms
on the "surface" to dimerize or the capping 
of the dangling bonds by adatoms.
From Fig.1, it can be seen that, as a result 
of reconstruction such as 
dimerization or capping, the bonding configurations 
associated with the exterior
atoms are distorted from the regular 
tetrahedral network, some more
so than others.

In Fig.2, the electronic density of states (EDOS) 
of the stable configuration
of Si$_N$ clusters for $N$=200, 800, and 2000 are 
shown, where the energy of the HOMO
for each case is taken to be zero. A key feature 
which stands out in all three plots of density of 
states is that there is 
no well-defined energy gap above the HOMO (E=0) 
in the band gap region of a semiconducting
Si system. Instead, defect
states due to the remnant dangling bonds and 
distorted bonding
configurations are found in the gap region. 
This calculation demonstrates that 
the disappearance of a clean gap is the price 
for achieving the stability of a Si nano-cluster 
on its own.
In other words, a "pure" and stable Si 
nano-cluster will not have a 
well-defined gap which is comparable to
the gap of a bulk Si.

\section {Role of Hydrogen Passivation on the 
Stability and the Existence of a Clean Gap for a Cluster 
with Complete Shell}

Previous theoretical studies on hydrogen-passivated 
{\it unrelaxed} Si clusters with
an overall symmetry and no local bond distortion have
shown the existence of a clean and well-defined energy 
gap and that the gap is blue-shifted
as the size of the nano-cluster is reduced \cite{SYR92}. 
However, the 
role of hydrogen passivation in eliminating the defect 
states have not been 
analyzed thoroughly in these studies. Also, a detailed
analysis of the stability of these
structures have not been provided. 

To shed light on these issues, we consider a Si$_{83}$ 
cluster whose atoms
occupy complete shells of a regular bulk tetrahedral 
network so that the cluster
not only possesses local tetrahedral bonding 
configuration but also an overall symmetry. 
The reason for choosing a small-sized cluster 
is that it is more convenient to
analyze its properties in terms of hydrogen 
passivation while the
underlying physics concerning hydrogen 
passivation is basically the same
regardless of the size of the system. In the following, 
we will investigate the role 
played by hydrogen passivation in eliminating the 
dangling bonds and stabilizing the system 
for both unrelaxed and relaxed clusters. In considering 
relaxed clusters, we will make 
further distinction between the case when the silicon 
cluster is first passivated and 
then relaxed and the one in which the cluster is first 
relaxed and then passivated.

We start by calculating the electronic DOS of 
the {\it unrelaxed} Si$_{83}$
cluster whose structure is identical to the 
bulk structure, 
including the bond length (as determined by 
Menon and Subbaswamy's 
NOTB Hamiltonian\cite{MM97}). For the unrelaxed 
Si$_{83}$ cluster, 
each of the 41 interior atoms has 4 bonds both 
by the simple distance 
criterion as well as by the bond charge 
criterion \cite{DRA99} 
while the remaining 42 exterior atoms have in 
total 108 dangling bonds. Clearly,
Si$_{83}$ with 108 dangling bonds is not a 
stable structure. The DOS for this structure 
is meant only to serve as a benchmark for 
comparisons with other cases. The result of this
calculation together with the density of 
states corresponding
to the cases of unrelaxed Si$_{83}$H$_{108}$ 
and relaxed Si$_{83}$ H$_{108}$ clusters,
respectively, are shown in Fig. 3. The 
unrelaxed structure of Si$_{83}$H$_{108}$ is obtained
by attaching a hydrogen atom for each dangling 
bond along the
tetrahedral bonding direction. For the hydrogen 
passivated Si clusters, we
fitted the NOTB Hamiltonian for Si-H bonding 
with respect to the bond
length and the molecular orbital energy of 
SiH$_4$ \cite{LC96,MVR95}. It should be noted 
that the DOS plots in this and subsequent 
figures are always displayed with the energy 
of HOMO for each case set to zero. 

A cursory examination of the DOS plots shown 
in Fig.3 reveals the following
points: (1) There is a well-defined and 
substantial clean gap for the 
DOS corresponding to the unrelaxed (unstable) 
Si$_{83}$ cluster. But this
gap does not begin at the energy of HOMO 
($E=0$). There are two prominent
peaks at the two extremities of the gap. 
The first peak is immediately
above the HOMO (E=0) while the second peak 
is above the gap. Comparing the appearance of a
well-defined gap in the DOS of this unrelaxed 
Si$_{83}$ with no
gap for relaxed clusters discussed in section 
II, one can easily 
associate the existence of a clean gap either 
with the local bonding configuration 
and/or the overall symmetry of the cluster. 
(2) With the passivation
of the 108 dangling bonds associated with the 
exterior atoms, the gap of 
the unrelaxed Si$_{83}$H$_{108}$ cluster 
"opens" up substantially with
respect to the gap of the unrelaxed Si$_{83}$. 
This gap now begins at 
the energy of HOMO. A comparison of these two 
density of states leads to an impression
that the "opening" of the gap in 
Si$_{83}$H$_{108}$ has resulted from the
elimination of the two broad peaks in the 
DOS of the unrelaxed Si$_{83}$,
one immediately above the energy of HOMO of 
Si$_{83}$ and the other across 
the gap from the first peak. Since there is 
no distortion in the bonding 
configurations of the exterior atoms in the 
unrelaxed Si$_{83}$, the defect 
states are entirely due to the 108 dangling 
bonds. One can then conclude 
that the opening-up of the gap is due to 
the saturation of the dangling bonds by hydrogen
passivation.
Furthermore, one might even be tempted to 
attribute the two broad peaks as 
associated with the dangling bond states. 
(3) After we relaxed the hydrogen
passivated Si$_{83}$H$_{108}$ cluster, we 
find that there is hardly any 
change in the major part of the gap region 
of the relaxed Si$_{83}$H$_{108}$ as
compared to that of the unrelaxed 
Si$_{83}$H$_{108}$, with only a slight
narrowing of the gap due to a small downward 
shift of the energy of LUMO
(lowest unoccupied molecular orbital).
The main features of the density of states 
of the two cases (unrelaxed and relaxed
Si$_{83}$H$_{108}$) are also very similar. 
There is, however, some modification
in the interior of the bands with the 
appearance of features which can be identified 
with Si-H bonding (see the discussion following 
figs. 7 and 8, respectively).

To build a complete understanding of the 
effects of passivation, we have
carried out a local analysis of the 
electronic structure of the unrelaxed
Si$_{83}$ as well as the unrelaxed 
Si$_{83}$H$_{108}$ clusters. In the 
framework of a NOTB approach, the electronic 
structure is determined by
solving a general eigenvalue problem
\begin{equation}
        H C_{\lambda} = E_{\lambda} S C_{\lambda} ,
\end{equation}
where $C_{\lambda}$ is the column 
vector representing the coefficients
of expansion of the wave function 
$\Psi _{\lambda}$ with the eigenenergy
$E_{\lambda}$ in terms of some 
localized basis set $\phi_{i\alpha}$ not 
explicitly stated. Here $i$ refers 
to the atomic site and $\alpha$ the 
orbitals at site $i$. The Hamiltonian 
matrix elements $H$ and the 
overlapping matrix elements $S$ are  
defined as
\begin{equation}
       H_{i\alpha,j\beta} = \int \phi_{i\alpha} (\vec{r})
                {\cal H} \phi_{j\beta} (\vec{r}) d\vec{r} ,
\end{equation}
\begin{equation}
       S_{i\alpha,j\beta} = \int \phi_{i\alpha} (\vec{r})
                 \phi_{j\beta} (\vec{r}) d\vec{r} .
\end{equation}
But in the semi-empirical NOTB approach, these 
matrix elements are treated as 
parameterized functions of 
$\vec{R}_{ij}=\vec{R}_j-\vec{R}_i$ with the
relevant parameters fitted to experimental 
results and/or first principles
calculations. The details of the NOTB 
formulation can be found in \cite{MM97}.

Within the framework of the NOTB treatment, 
the electronic DOS can be 
expressed as \cite{CSJ98,DRA99}
\begin{eqnarray}
  \rho (E) & = & \sum_{\lambda, i\alpha, j\beta} c_{i\alpha,\lambda}
        c_{j\beta,\lambda} S_{j\beta,i\alpha} \delta(E-E_{\lambda}) 
         \nonumber \\
      &=& \lim _{\epsilon \rightarrow 0} \frac{1}{\pi}
         \sum_{\lambda, i\alpha, j\beta} c_{i\alpha} c_{j\beta}
        S_{i\alpha,j\beta}\frac{\epsilon}{(E-E_\lambda)^2+\epsilon^2}
\end{eqnarray}
Equation (4) allows one to rewrite the DOS in terms of the local DOS (LDOS)
$\rho_i(E)$ such that
\begin{equation}
    \rho(E) = \sum_i \rho_i(E)
\end{equation}
with
\begin{equation}
  \rho_i(E) = \sum_{\lambda,\alpha,j\beta}c_{i\alpha}c_{j\beta}
        S_{j\beta,i\alpha} \delta (E-E_\lambda)
\end{equation}

The NOTB Hamiltonian used in this study was developed in a $sp^3$ framework.
For the unrelaxed Si$_{83}$ cluster, there are then 83x4=332 possible
states. Subtracting the 108 dangling bond states, there are 224 states with 
half of them (112) being the bonding states and the other half anti-bonding
states. In the meantime, there are 332 valence electrons in Si$_{83}$ and they 
occupy the lowest 166 states. Out of these 166 states, 112 are the bonding 
states. This means that the remaining 166-112=54 states must come from the 108 
dangling bond states, leaving the other 54 dangling bond states unoccupied.
To understand the characteristics of these states, we define a quantity
\begin{equation}
     A_i(E_{\lambda_1},E_{\lambda_2}) = \int_{E^-_{\lambda_1}}^{E^+_{\lambda_2}}
          \rho_i(E) dE,
\end{equation}
where $E^-_{\lambda_1}$= $E_{\lambda_1}- 0^+$ and
$E^+_{\lambda_2}$=$E_{\lambda_2}+ 0^+$.
This quantity is the local contributions to the integrated DOS for the 
eigenstates in the energy interval $[E_{\lambda_1}$,$E_{\lambda_2}]$. 
Hence this
quantity will be referred as the partially integrated LDOS (PILDOS).
When PILDOS is plotted against the site index $i$, it provides a transparent 
picture as to
whether states in a certain energy interval are extended through out the 
cluster or only restricted in a certain region of the cluster. (In this work,
atoms are labeled from inside to outside.) In Fig.4,
$A_i$ for the following energy intervals are shown, namely $A_i(E_1,E_{112})$,
$A_i(E_{113},E_{166})$, and $A_i(E_{167},E_{220})$. It can be seen that the 
$A_i$ for the first 112 states, $i.e.$ $A_i(E_1,E_{112})$ is almost uniformly 
distributed through out the 83 sites of the unrelaxed Si$_{83}$ cluster.
Hence the first 112 states must be the 112 extended bonding states. The $A_i$
corresponding to the next 54 states, $A_i(E_{113},E_{166})$ is mostly
concentrated in the region of the 42 exterior sites, indicating that these
54 states (113 to 166) must be the occupied dangling bond states. The
$A_i$ for the 54 states from 167 to 220 is almost entirely localized in 
the "surface" region of the 42 exterior sites. Hence these 54 states must
correspond to the 54 unoccupied dangling bond states. In Fig.5, the
$A_i$'s for the 54 states from 221 to 274 are shown in the interval of
every 18 states, namely, $A_i(E_{221},E_{238})$, $A_i(E_{239},E_{256})$, and
$A_i(E_{257},E_{274})$.  It can be seen that these $A_i$'s are distributed 
through out the 83 sites of the cluster, indicating that these 54 states
must be the extended anti-bonding states. Thus the first broad peak in the DOS
of the unrelaxed Si$_{83}$ just above the energy of HOMO (seen in Fig.3)
which covers the energy range from $E_{167}$ to $E_{220}$ is indeed 
composed of the dangling bond states (unoccupied) while the second broad
peak above the gap, covering the energy range from
$E_{221}$ to $E_{274}$, is composed entirely of extended anti-bonding states.

For the unrelaxed Si$_{83}$H$_{108}$ where the 108 dangling bonds of 
Si$_{83}$ are saturated by the 108 hydrogen atoms, there are all
together 83x4+108x1=440 states with the 220 lowest states being the
bonding states and the remaining 220 states anti-bonding states.
There are also 440 electrons in this cluster. Hence the 220 bonding
states are completely occupied while the 220 anti-bonding states are
unoccupied. The bonding and anti-bonding states are separated by
a wide band gap. The picture presented above can be understood in terms
of the $A_i$'s for the unrelaxed Si$_{83}$H$_{108}$ shown in Fig.6
where $A_i(E_{113},E_{220})$ and $A_i(E_{221},E_{328})$ are plotted.
It can be seen that both $A_i$'s are distributed uniformly through out
the 83 Si sites of the cluster with only very minor contributions
from the outer 108 hydrogen sites, indicating that (i) these states are
extended throughout the Si cluster,and (ii) the widening of the
gap is due to elimination of the localized dangling bond states.
In Figs.7
and 8, the LDOS's at two typical hydrogen sites 
corresponding to the elimination
of two dangling bonds of a Si atom and those at three typical hydrogen
sites corresponding to elimination of three dangling bonds of a Si
atom are shown respectively. It can be seen that there is absolutely no
contribution from the hydrogen site to the band edge states of the DOS of the 
unrelaxed Si$_{83}$H$_{108}$ cluster. The combined picture of Figs. 6, 7, 
and 8 shows convincingly that the existence of the clean band gap in
the DOS of the unrelaxed Si$_{83}$H$_{108}$ is entirely due to the
Si atoms in the cluster with no contribution from the surface chemistry
of Si-H bonding. In other words, the role of Hydrogen is simply to eliminate
the dangling bonds at the surface of Si atoms.

We have also carried out the relaxation of Si$_{83}$ cluster to obtain a
stable Si$_{83}$ structure. The relaxation process indeed eliminated many
of the 108 dangling bonds associated with the unrelaxed Si$_{83}$. But this occurs 
at the expense of introducing severe bond distortion for the "surface" atoms,
resulting in the appearance of bond distorted defect states in the gap (see Fig. 9).
The coexistence of the distorted bond states and the remnant dangling bond states that are not
eliminated by relaxation provides the explanation why Si nanoparticles deposited onto a film and aged
in vacuum show no Photoluminescence \cite{seraphin}. The relaxed stable Si$_{83}$ structure 
(Si$^{r}_{83}$) has only 22 dangling bonds. We then
saturated these 22 dangling bonds by hydrogen and 
relaxed the Si$^{r}_{83}$H$_{22}$
configuration. In Fig.9, the density of states corresponding to 
unrelaxed Si$_{83}$, relaxed Si$_{83}$, and relaxed 
Si$^{r}_{83}$H$_{22}$ clusters are 
shown. It can be seen that the relaxation of Si$_{83}$ eliminated the
majority of the dangling bonds and the passivation by hydrogen after the 
relaxation further reduced the remaining dangling bonds. But the distorted
bonding configurations resulting from the "surface" reconstruction are mostly 
still present even for relaxed Si$^{r}_{83}$ H$_{22}$. Hence 
hydrogen passivation of relaxed cluster is not expected to yield the 
best clean gap.

Finally, in Table I, we listed the comparison of the total energy for all
the cases considered. It can be seen that the relaxed Si$_{83}$H$_{108}$ has
the lowest energy (-5429eV) while the unrelaxed Si$_{83}$H$_{108}$ has almost
the same energy (-5419eV). The two structures are also very similar,
indicating that hydrogen passivation before the cluster has the time to relax 
is the best way to fabricate stable clusters with a clean gap.

\section{Hydrogen passivation of clusters with incomplete shell}

As pointed out in the introduction, it is very difficult to control 
experimentally the number of atoms in a cluster. This simply means
that, in the fabrication process, there is no way to be certain that
the cluster will have exactly the desired number of atoms to complete the shell
of, say, a tetrahedral network. The question must then be raised as 
to whether a stable hydrogen-passivated cluster possessing a clean
gap can be fabricated. To shed light on this situation, we have considered
the hydrogen passivation of the unrelaxed cluster Si$_{200}$. The unrelaxed
Si$_{200}$ cluster is constructed in the following way: The Si atoms are
placed on the regular tetrahedral sites. But on the last incomplete
shell, the Si atoms are distributed on the tetrahedral site in a random
manner. Hence a cluster constructed in this way will have undistorted local
bonding configuration but not an overall symmetry. 
For the particular cluster considered,
there are 158 dangling bonds associated with this unrelaxed configuration.
We passivate these dangling bonds with hydrogen atoms and calculate the
DOS of both unrelaxed Si$_{200}$H$_{158}$ and 
relaxed Si$_{200}$H$_{158}$ clusters.
In Fig. 10, the density of states for the unrelaxed Si$_{200}$, unrelaxed
Si$_{200}$H$_{158}$ and relaxed Si$_{200}$H$_{158}$ are shown. 
It can be seen that the unrelaxed Si$_{200}$ cluster, as
in the case of a cluster with filled shells (eg. Si$_{83}$),
possesses a clean gap and the lower edge of this gap does not
begin at HOMO. The unrelaxed but hydrogen passivated 
Si$_{200}$H$_{158}$ has a much wider gap than that of 
the unrelaxed Si$_{200}$ 
and this gap begins at HOMO. Finally, there is hardly
any change in the width of the gap between the relaxed and unrelaxed 
Si$_{200}$H$_{158}$ clusters. We find that results for a cluster with
an incomplete shell are analogous to those for a cluster 
with filled shells (eg. Si$_{83}$)
The fact that the unrelaxed Si$_{200}$ possesses a 
clean gap indicates that
the existence of the gap only depends on the local 
bonding configuration rather 
than the overall symmetry.

The findings shown in Fig.10 indicate that one may produce a 
cluster with a well-defined clean gap by passivating an unrelaxed
cluster with hydrogen regardless of the size of the cluster (i.e. number
of atoms in the cluster). But, the passivation must be completed before 
the cluster has the time or the opportunity to relax. Detailed information
on the time scales in the above-mentioned process requires a
study of the kinetics of hydrogen passivation. In the absence of such detailed calculations, 
one can gather evidences from experiments and make some general remarks. It is known
that hydrogen passivation of Si(100) surface at 295 K leads to the formation
of 1x1 surface \cite{Terashi}. Carrying over this analogy to the cluster case, one
may speculate that it should be possible to hydrogen passivate an unrelaxed silicon cluster 
around room temperature. 

\section{The size dependence of the gap for a general cluster}

In this section, we study the size dependence of
the energy gap for relaxed, hydrogen-passivated Si clusters of
arbitrary sizes with no bond distortion. Relaxation 
ensures that the cluster is in a 
stable equilibrium configuration.
Since there is no longer an overall symmetry 
associated with such a general cluster, no symmetry-based 
consideration can be used to reduce the computing effort of
solving the eigenvalue problem and/or determining the DOS
(see, for example, Ref. \cite{SYR92}). For such a general 
cluster with a size beyond
1000 atoms, the calculation of DOS becomes computationally
excessive. While the method of real space Green's function
(RSGF) is efficient and convenient to calculate the DOS for 
large systems with no or reduced symmetry \cite{SYW95}, 
it is not a convenient tool to determine the energy gap, 
particularly if one wants to achieve good accuracy for
the gap. This is because an accurate determination of the gap
requires an accurate calculation of the energies of both
the HOMO as well as the LUMO. To do this with the method of
RSGF, one has to calculate $Im G(E+i\epsilon)$ in the regions
of the band edges at an energy interval sufficiently small
compared to $\Delta E_{av}$, the average separation between
consecutive eigenvalues. In addition, one has to calculate
these quantities for a series of decreasing $\epsilon $
(smaller than $\Delta E_{av}$) to assure the convergence
of the result. The whole process, while doable, can be time
consuming. Therefore, we have developed a scheme, 
as described in the Appendix I,
that directly computes the eigenvalues corresponding to HOMO and LUMO 
without having to obtain the entire eigenvalue spectrum. This method
is ideally suited for situations where selected 
eigenvalues of a large matrix are
required.

Using the method outlined in the Appendix, we have 
carried out a series 
of very accurate calculation of the energy gap for 
relaxed, hydrogen-passivated $Si_NH_M$ clusters
of various sizes with no bond distortion. In particular, 
we have performed the energy gap calculation
for the following sets of N (number of Si atoms) 
and M (number of H atoms), respectively:
$(N,M)$ = (200,158),(800,372),(1400,518),
(2000,636),(3000,828),(4000,1012), (5000,1276),
and (6000,1354).
Denoting the energy gap of a cluster of diameter 
$d =3.3685 N^{1/3}$ (in \AA , see Ref. \cite{AZ96})
as $E_g(d)$ and that of the bulk as $E_{g0}$ (both in eV), 
we have shown in Fig. 11, the log-log plot of 
$\frac{E_g(d)}{E_{g0}}-1$ as a function of $d$.

We find that the energy gap as a function
of diameter of the cluster can be fitted to an equation of the form

\begin{equation}
  \frac{E_g(d)}{E_{g0}}- 1 = \frac{24.98}{d^{1.364}}  . 
\end{equation}
We compare this equation with the equation for the size dependence
of the energy gap of hydrogen-passivated silicon clusters with
complete shells as given in Ref. \cite{AZ96}, namely, 

\begin{equation}
   \frac{E_g(d)}{E_{g0}}-1 = \frac{88.34/E_{g0}}{d^{1.37}}. 
\end{equation}
Since the energy gap of bulk Si obtained 
with the NOTB Hamiltonian \cite{MM97} used in this calculation
has a value $E_{g0}=3.16eV$, one can rewrite Eq. (9) to yield 

\begin{equation}
  \frac{E_g(d)}{E_{g0}}-1= \frac{27.96}{d^{1.37}}  .
\end{equation}
The comparison of Eqs.(8) and (10) indicates that 
for the relaxed, hydrogen-passivated Si nano-cluster of arbitrary size 
and with no bond distortions, the size-dependence of its energy gap follows 
almost an identical pattern as that for the hydrogen-passivated clusters
with full symmetry and no bond distortions.

\section{Summary}

We have identified the factors responsible for 
the stability and the origin of a clean energy 
gap of a silicon nanocluster. We find that the 
existence of local configurations with no bond distortion
is the key to the emergence of an energy gap. 
This is demonstrated by considering
unrelaxed clusters with no bond distortion for 
which an energy gap exists no matter whether we 
consider a cluster with full shell (i.e. overall 
symmetry) or a cluster with an incomplete shell. However,
the energy gap of unrelaxed clusters does not 
start at HOMO. This is due 
to the presence of dangling bond states. We have 
shown that the localized dangling bond states are 
completely eliminated by hydrogen passivation, 
resulting in a wider and clean energy gap. 
Although the relaxation process can also eliminate 
dangling bond states, we find that this process
introduces impurity states in the gap associated 
with the bond distortion of surface
atoms in the cluster. These impurity modes can not 
be eliminated by further saturation of remnant
dangling bonds with hydrogen. In conclusion, we 
find that the relaxed hydrogen-passivated 
silicon nanoclusters with no bond distortions 
have the cleanest energy gap and the most stable 
structure among the different cases studied in this paper.

\noindent {\bf Acknowledgment}

This work was supported by the NSF grant DMR-9802274. 

\newpage
\noindent {\bf Appendix I.  A Method For Calculating 
the Selected Eigenvalues of a Large Matrix}

This method for finding selected 
eigenvalues is built around the
ideas of real-space Green's function \cite{SYW95} 
and the conjugate gradient 
method \cite{WHP86}. 

Starting from a generalized eigenvalue equation 
given in Eq. (1), it can be shown easily that
$$
    G(E)SC_{\lambda} = \frac{1}{E-E_{\lambda}}C_{\lambda} ,
\eqno {(A1)}
$$
where $G(E)$ is the (generalized) Green's function
defined by
$$
    G(E) = (ES-H)^{-1} .
\eqno {(A2)}
$$
To determine a particular eigenvalue $E_{\lambda_0}$ 
of the matrix,  
we start with an arbitrary vector $C^{(0)}$ and 
then generate a series of vectors $C^{(n)}$ 
iteratively according to the equation
$$
    C^{(n)} = G(E)SC^{(n-1)},\; \; \; n=1,2,3,\cdots .
\eqno {(A3)} 
$$
By writing $C^{(0)}$ as a
linear combination of eigenvectors $C_\lambda$, namely, 
$$
    C^{(0)} = \sum_\lambda a_\lambda C_\lambda ,
\eqno {(A4)}
$$
it can be seen that
$$
    C^{(n)} = \sum_\lambda \frac{a_\lambda}{(E-E_{\lambda})^n}C_{\lambda} .
\eqno {(A5)}
$$
By choosing the value of $E$ to lie close to 
$E_{\lambda_0}$, a particular eigenvalue among 
all the eigenvalues of the matrix, it
is possible to determine the value of $E_{\lambda_0}$ 
as accurately as one desires using the formula
$$
    E_{\lambda_0} = E - \lim_{n \rightarrow \infty }
            \frac{C^{(0)~T}SC^{(n)}}{C^{(0)~T}SC^{(n+1)}} .
\eqno {(A6)}
$$

In principle, this method can be used to calculate any particular
eigenvalue and its corresponding eigenvector for large systems
if we take advantage of the real space Green's 
function (RSGF)\cite{SYW95} method for calculating $G(E)$.
However, the procedure as outlined in Eqs.(A3) and (A6) requires
the repeated multiplication of the matrix $G(E)$ on a vector.
Since $G(E)$ is not a sparse matrix even when $H$ and $S$ 
are sparse matrices, the computational effort and computing resources needed
for calculating the series
of $C^{(n)}$ can become excessive when the system size under consideration
is very large and the computing resources are limited to desktop workstations. 
Therefore, we have developed an alternative scheme which takes
advantage of the sparseness of the matrices $H$ and $S$. 
In this scheme, we avoid
the step involving the calculation of the Green's 
function. We invert the
expression given in Eq. (A3) such that we can now 
determine $C^{(n)}$ from $C^{(n-1)}$ by
solving the equation
$$
    (ES-H)C^{(n)} = SC^{(n-1)} ,
\eqno {(A7)}
$$
using the conjugate gradient method as 
outlined in Ref.\cite{WHP86}. Once consecutive 
$C^{(n)}$'s are determined, they can be substituted 
into Eq.(A6) to determine $E_{\lambda_0}$.

In implementing the conjugate gradient scheme, 
we start from an arbitrary vector
$C^{(n)}_0$ and choose $A_0$ and $B_0$ such that
 
$$ 
A_{0} = B_{0}= SC^{(n-1)}-(ES-H)C^{(n)}_0
\eqno {(A8)}
$$
from which a sequence of vectors $A_i$, $B_i$, and $C^{(n)}_i$
$(n=1,2,3,\dots)$ can be generated by using the following set of formulae:
$$
    A_{i+1} = A_{i} - \alpha_i (ES-H) B_i ,
\eqno {(A9)}
$$
$$
    B_{i+1} = A_{i+1} + \beta_i  B_i ,
\eqno {(A10)}
$$
$$
    C^{(n)}_{i+1} = C^{(n)}_{i} + \alpha_i B_i ,
\eqno {(A11)}
$$
where 
$$
    \alpha_i = \frac{A_{i}^{~T} A_{i}}{B_i^{~T}(ES-H)B_i} ,
\eqno {(A12)}
$$
$$
    \beta_i = \frac{A_{i+1}^{~T} A_{i+1}}{A_i^{~T}A_i} .
\eqno {(A13)}
$$

The iterative procedure can be terminated
when $A_{m} \rightarrow 0$ and when this condition is satisfied, it can 
be shown that $C_{m}^{(n)} \rightarrow C^{(n)}$. This is proved below.
Using Eqs. (A8), (A9) and (A11), it can be shown that 

$$
\begin{array}{rcl}
   (ES-H)C_m^{(n)} & = & (ES-H)C_m^{(n)}+A_m \nonumber \\
                   & = & (ES-H)C_{m-1}^{(n)}+A_{m-1} \nonumber \\
                   & = & \cdots   \nonumber \\
                   & = & (ES-H)C_{0}^{(n)}+A_{0}  \nonumber \\
                   & = & SC^{(n-1)}  .  \nonumber \\
\end{array}
\eqno {(A14)}
$$
Using  $SC^{n-1}$ as given in Eq. (A7) in the above equation, it can be
seen that indeed 
$$
    C_m^{(n)}\rightarrow C^{(n)}
\eqno {(A15)}
$$
when $A_{m} \rightarrow 0$.

The main computational burden in calculating the $C^{(n)}$ and hence the
eigenvalue as given by Eq. (A6) is now shifted to the 
calculation $(ES-H)B_i$, which involves the multiplication 
of a matrix and a vector. 
However, this matrix is sparse because both $S$ and $H$ 
for most systems of interest are sparse in the real-space 
representation.
Because of this reason, the
computational effort in evaluating specific eigenvalues becomes
quite tractable even when the dimension of the system 
matrix is large.  In the present
work, the largest matrix that we have considered has 
the dimension 25354x25354 and
the computation was done on a HP J-282 (180 MHZ) with 256 MB RAM.

\newpage

\noindent{\bf FIGURES}

\vskip 0.1in

\noindent {FIG. 1 Optimized Structure of $Si_{2000}$ 
as obtained by the O(N)/NOTB-MD method.}
\vskip 0.1in
\noindent {FIG. 2  A comparison of the normalized 
electronic density of states of relaxed $Si_{200}$,
$Si_{800}$ and $Si_{2000}$, respectively. 
No discernible gap is seen in this
case. In this figure, the energy of HOMO level is fixed at zero.} 
\vskip 0.1in
\noindent{FIG.3  A comparison of the normalized electronic 
density of states of the unrelaxed $Si_{83}$
unrelaxed $Si_{83} H_{108}$, and relaxed $Si_{83} H_{108}$, 
respectively. The energy
of HOMO for each case is set at zero.}
\vskip 0.1in
\noindent{FIG.4  Partially integrated local density of states 
is plotted for three different 
energy intervals (two below and one above the HOMO level) at 
the sites of the unrelaxed 
$Si_{83}$ cluster.}
\vskip 0.1 in
\noindent{FIG.5  Partially integrated local density of states 
is plotted at the sites of the 
unrelaxed $Si_{83}$ cluster for energy eigenvalue index 
221 to 274 in steps of 18.}
\vskip 0.1 in
\noindent{FIG.6 Partially integrated local density of 
states is plotted at the sites of
unrelaxed $Si_{83} H_{108}$ cluster. }
\vskip 0.1 in
\noindent {Fig. 7
Local density of states at hydrogen sites which typically saturate 
two dangling bonds associated with an exterior Si atom.}
\vskip 0.1 in
\noindent {Fig. 8 
Local density of states at hydrogen sites which typically saturate 
three dangling bonds associated with an exterior Si atom.}
\vskip 0.1 in
\noindent {Fig. 9 
A comparison of the normalized electronic density of 
states of unrelaxed $Si_{83}$, 
relaxed $Si_{83}$, and relaxed $Si_{83} H_{22}$ whose 
remnant dangling bonds are
saturated with hydrogen atoms, respectively.}
\vskip 0.1 in
\noindent {Fig. 10
A comparison of the normalized electronic density of 
states of unrelaxed $Si_{200}$, 
unrelaxed $Si_{200}H_{158}$, and relaxed $Si_{200}H_{158}$.}
\vskip 0.1 in
\noindent {Fig. 11 The energy gap of the hydrogen-passivated
silicon cluster as a function of the size of the cluster.}

\newpage
\begin{table}
\caption{A comparison of the total energy for the 
various cases of Si$_{83}$ clusters 
studied in this work} 
\end{table}
\begin{center}
\begin{tabular}{|c|c|c|c|c|} \hline\hline 
\multicolumn{2} {|c|} {Si$_{83}$} &\multicolumn{2} 
{c|}{Si$_{83}$H$_{108}$}&Si${^r}_{83}$H$_{22}$\\  \hline
Unrelaxed & Relaxed & Unrelaxed & Relaxed &Relaxed\\  \hline
-3724eV & -3785eV &-5419eV &-5429eV &-4113eV \\  \hline
\end{tabular}
\end{center}

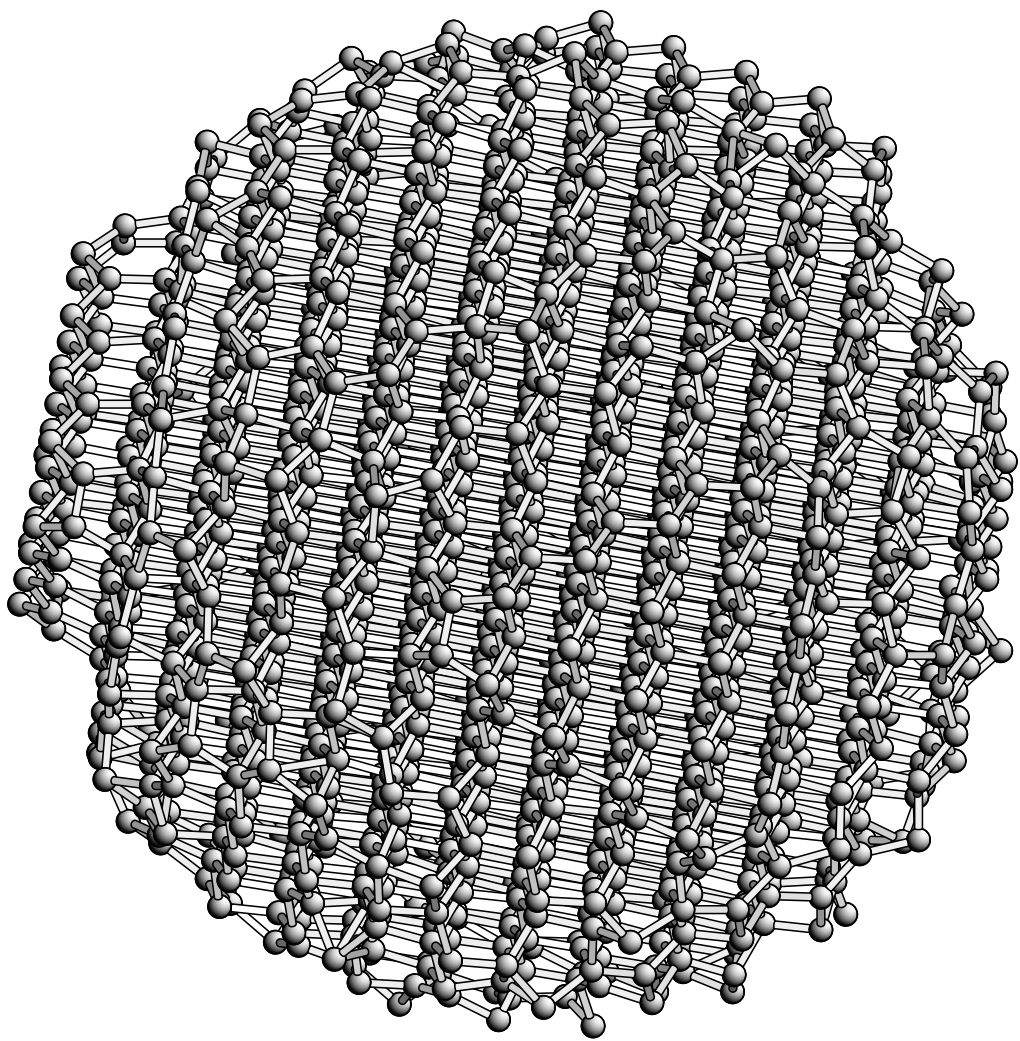
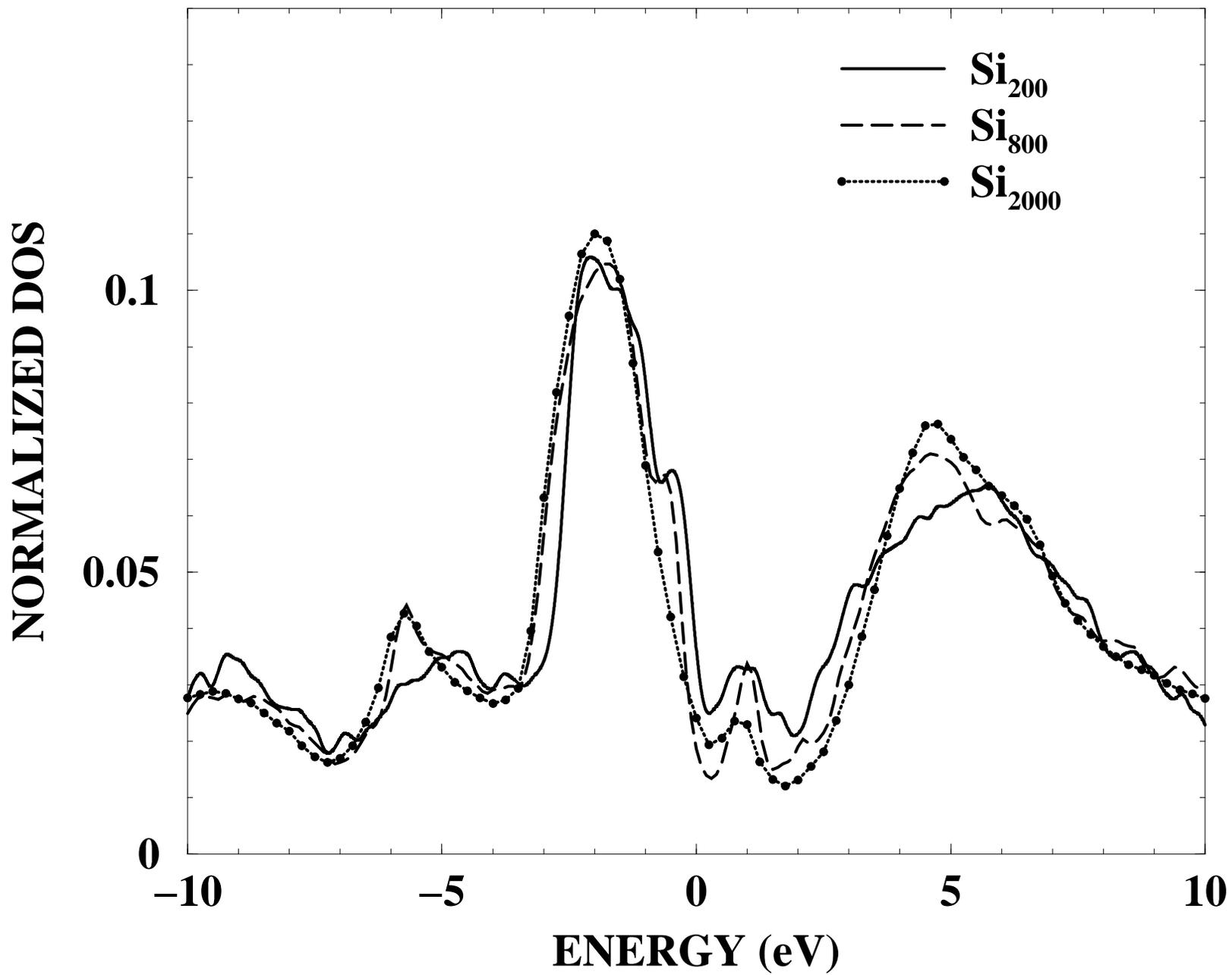
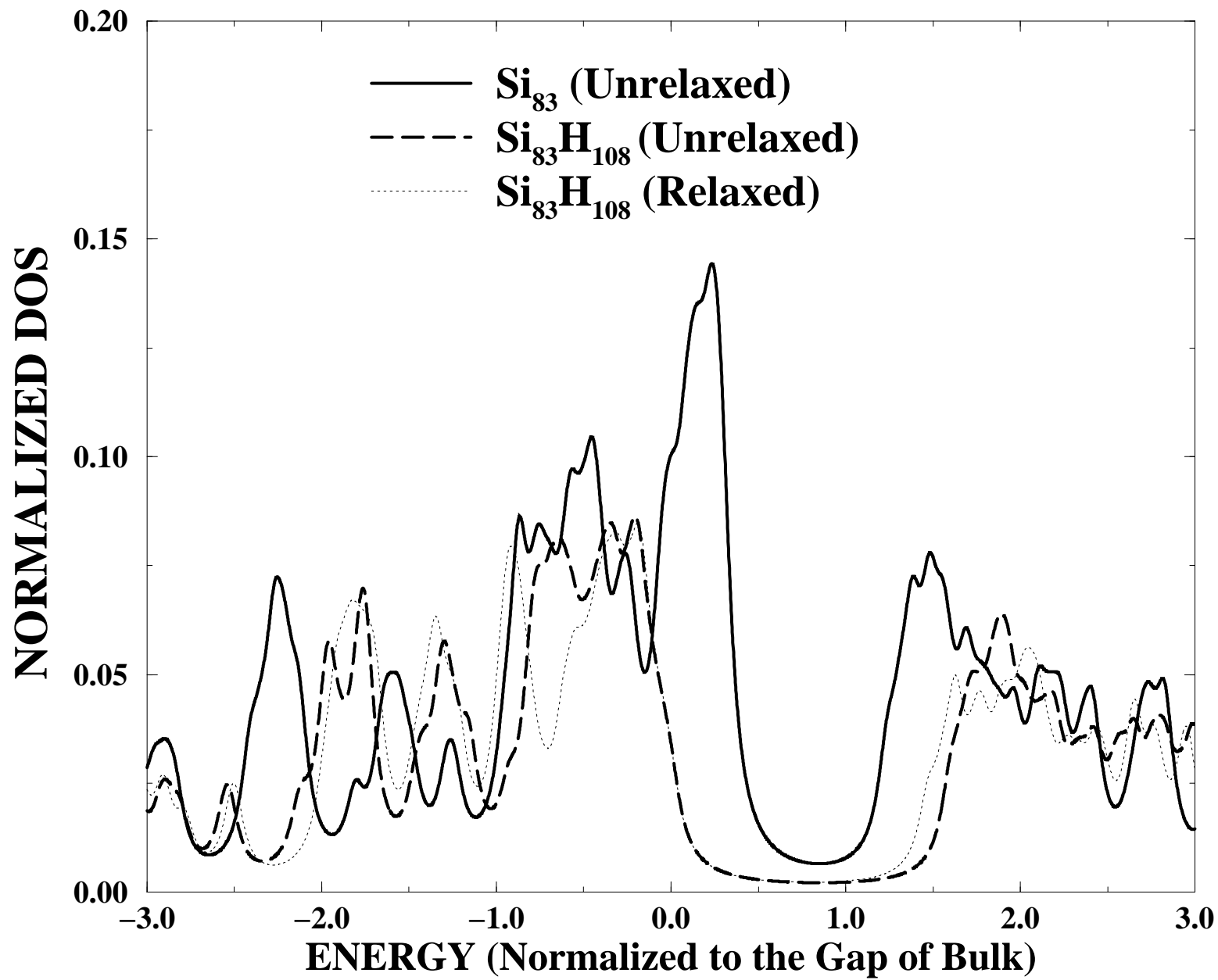
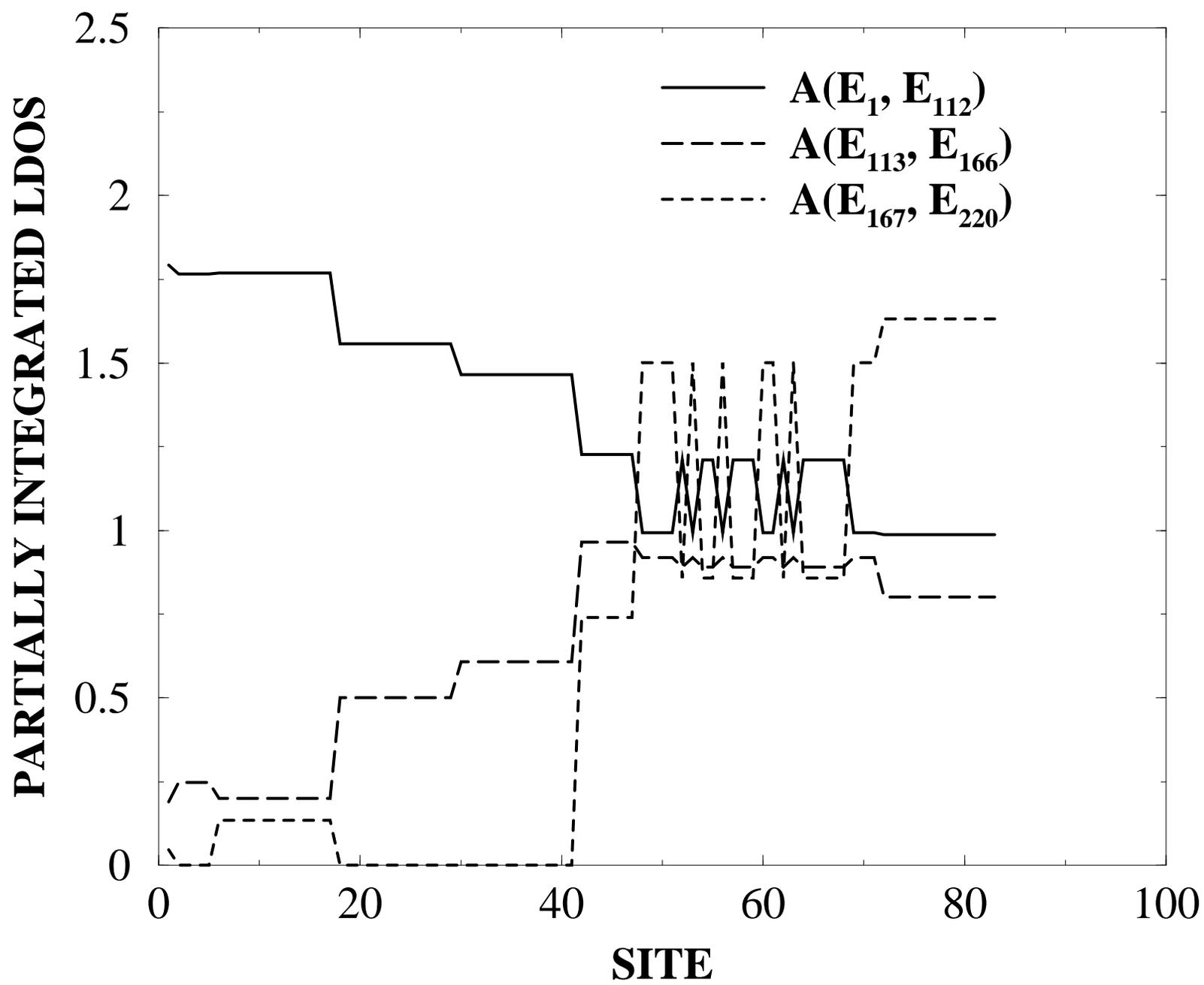
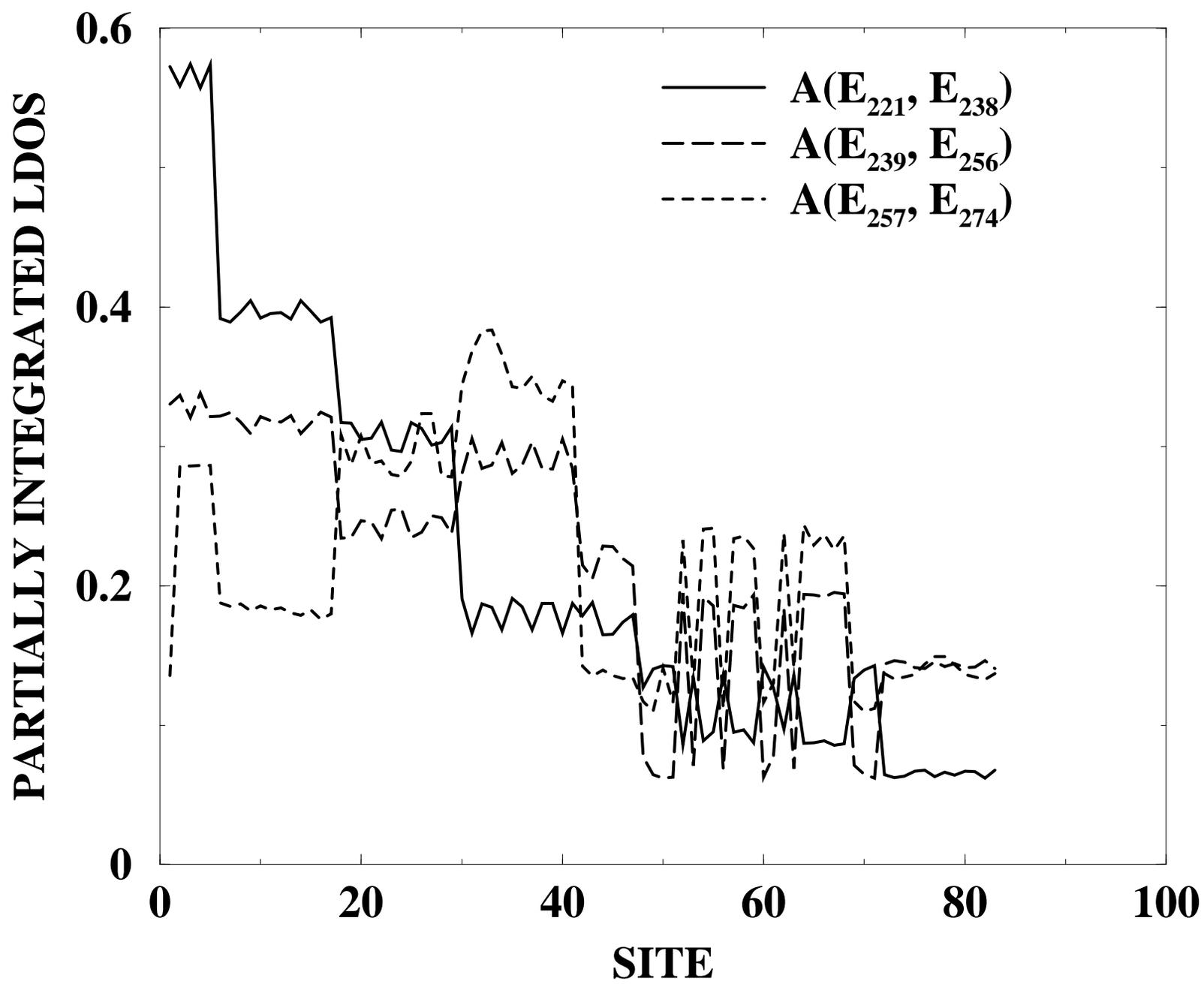
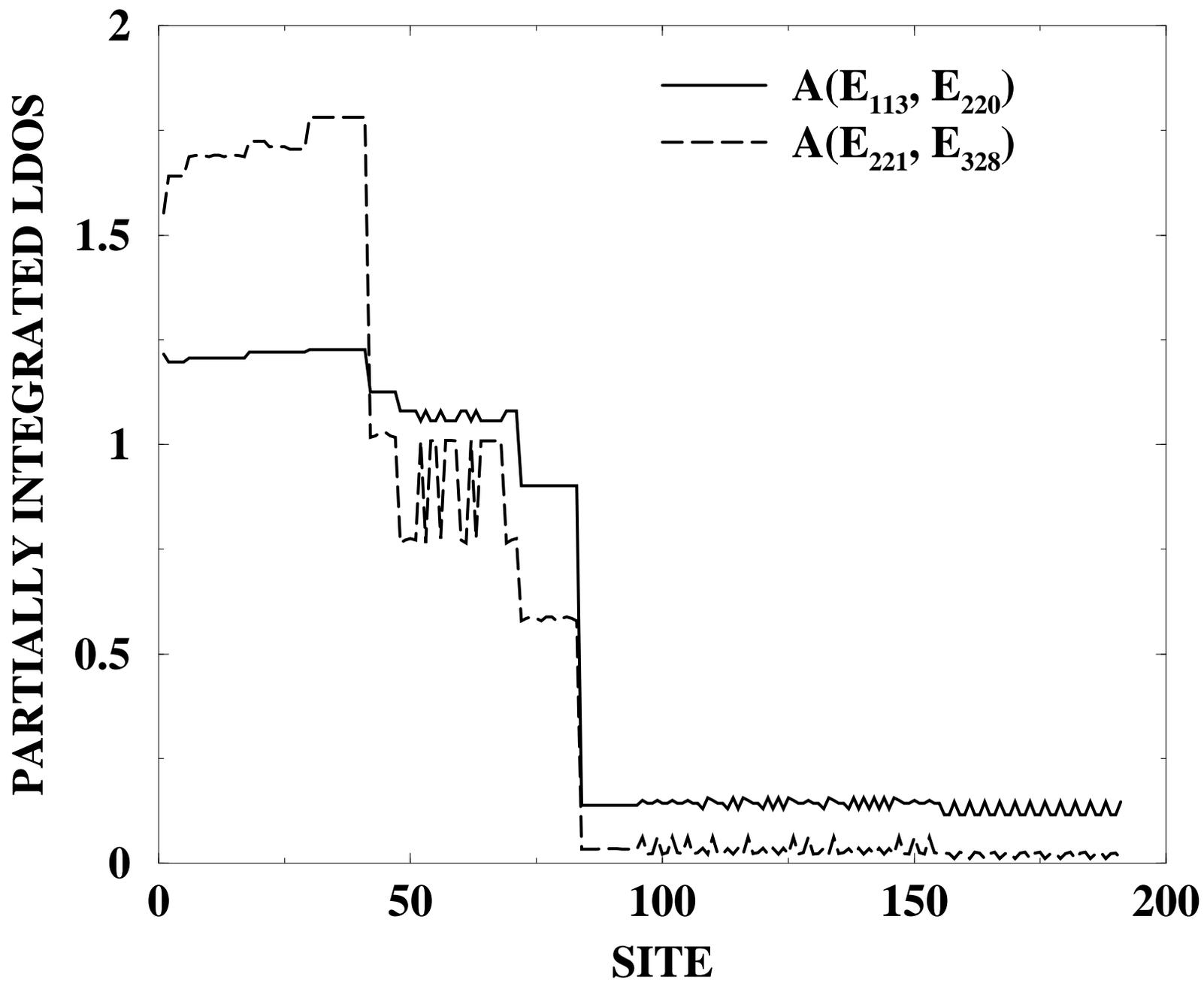
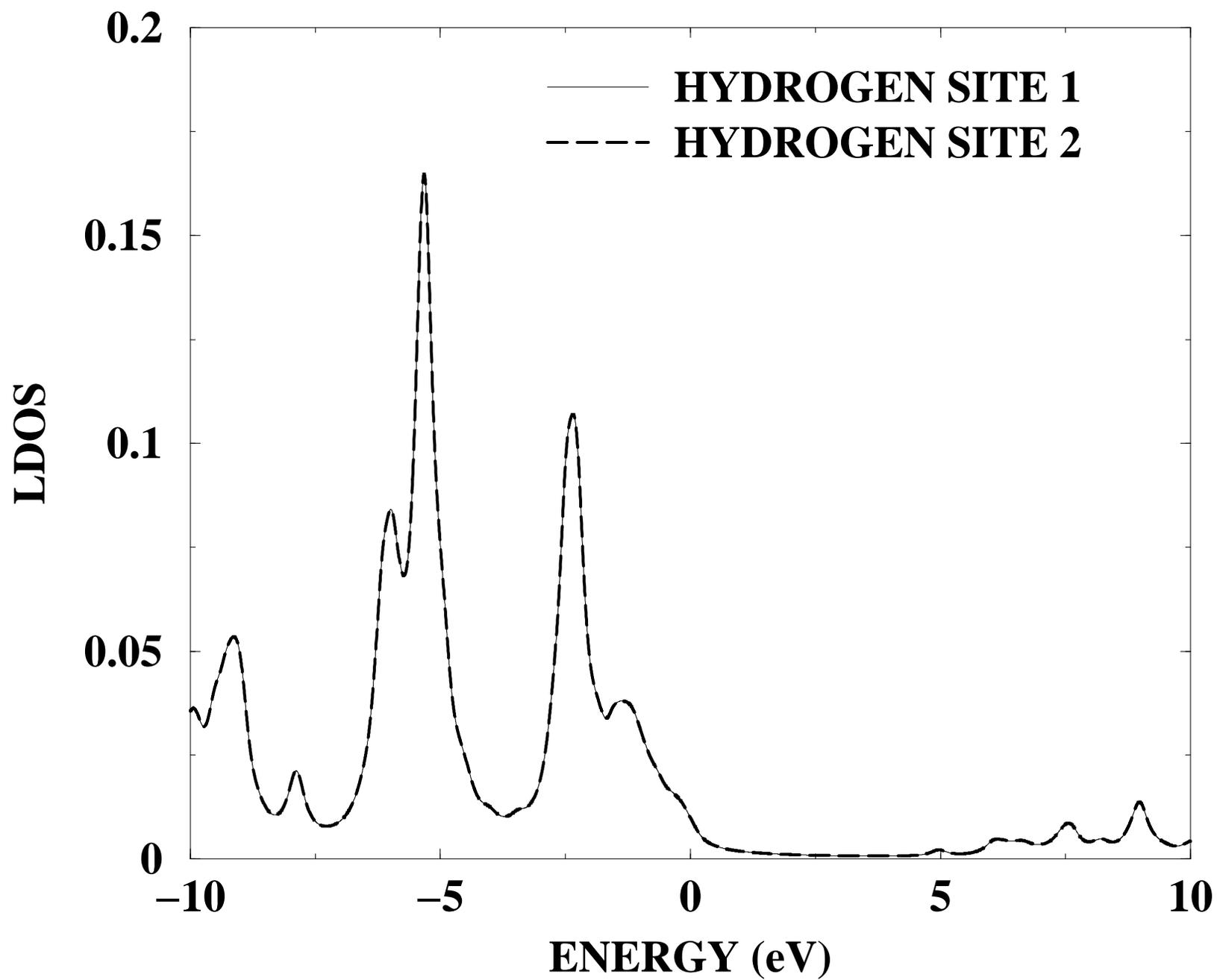
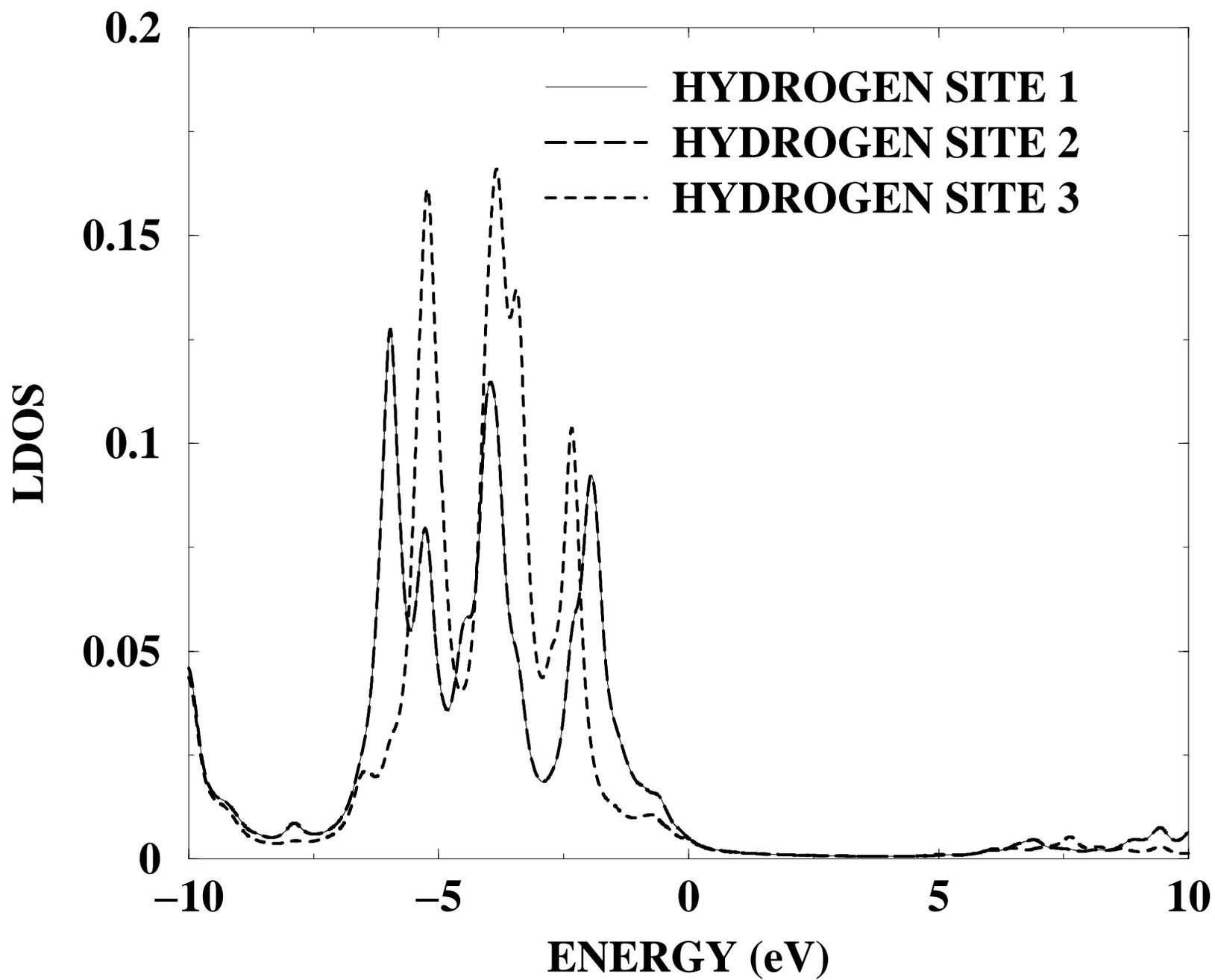
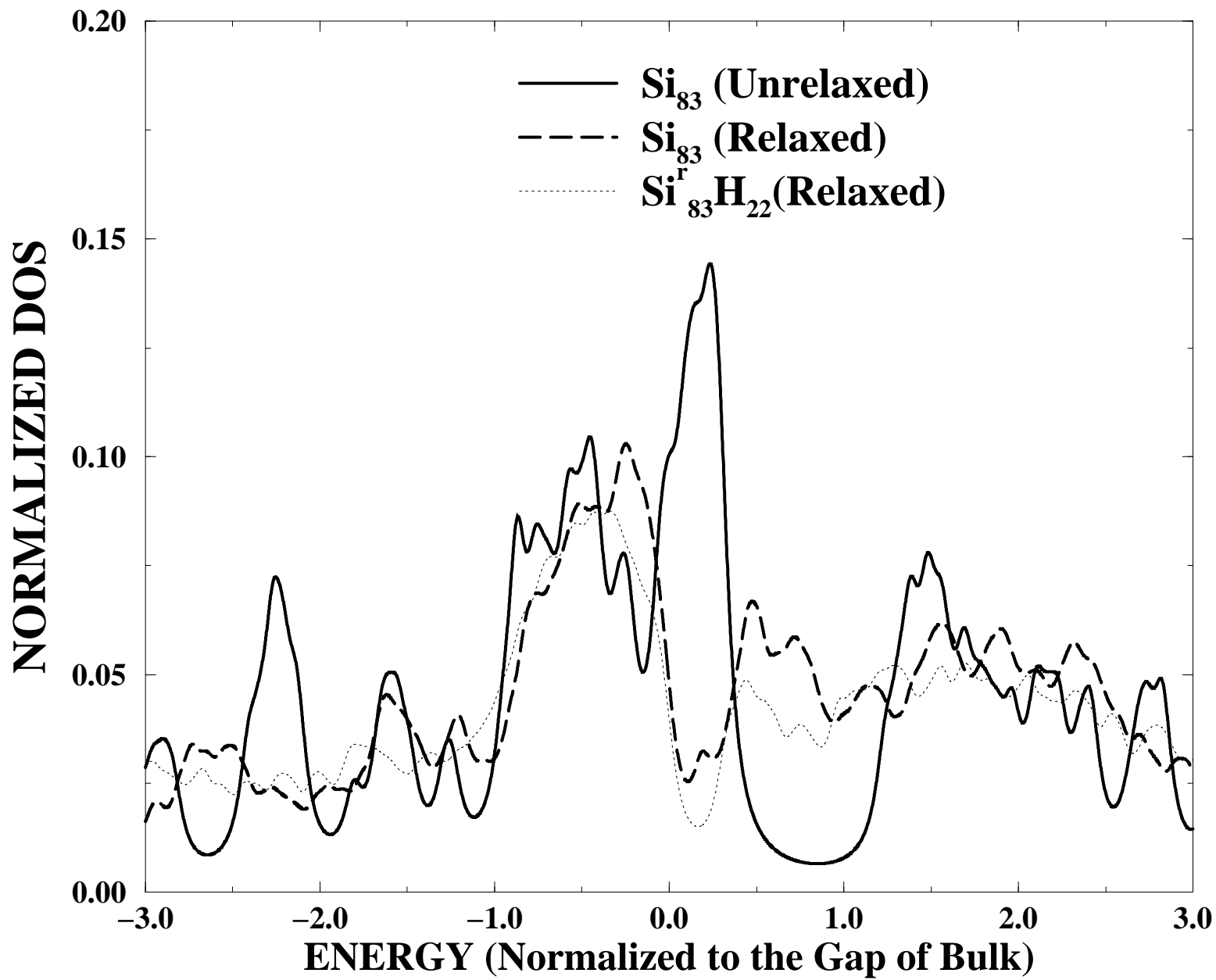
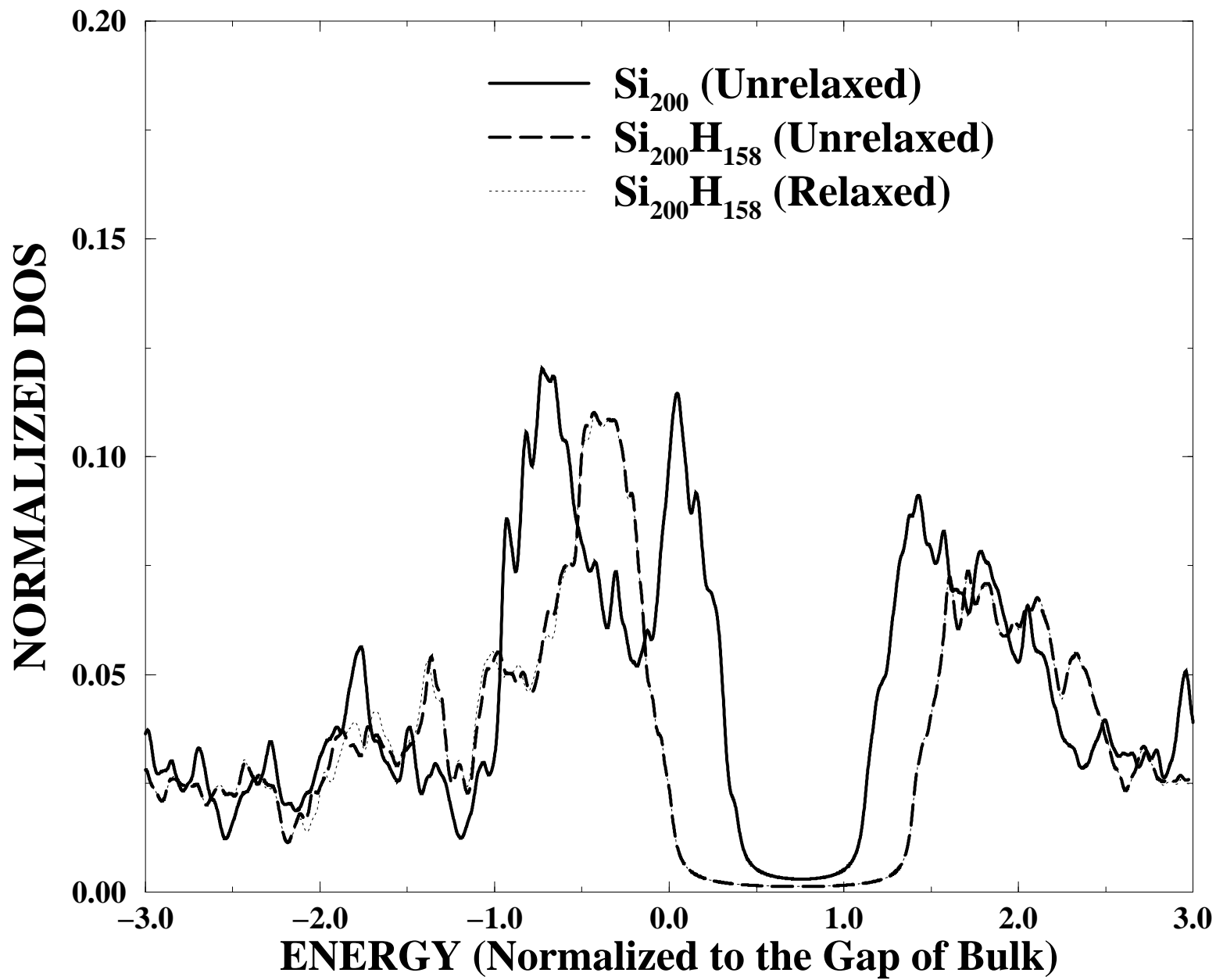
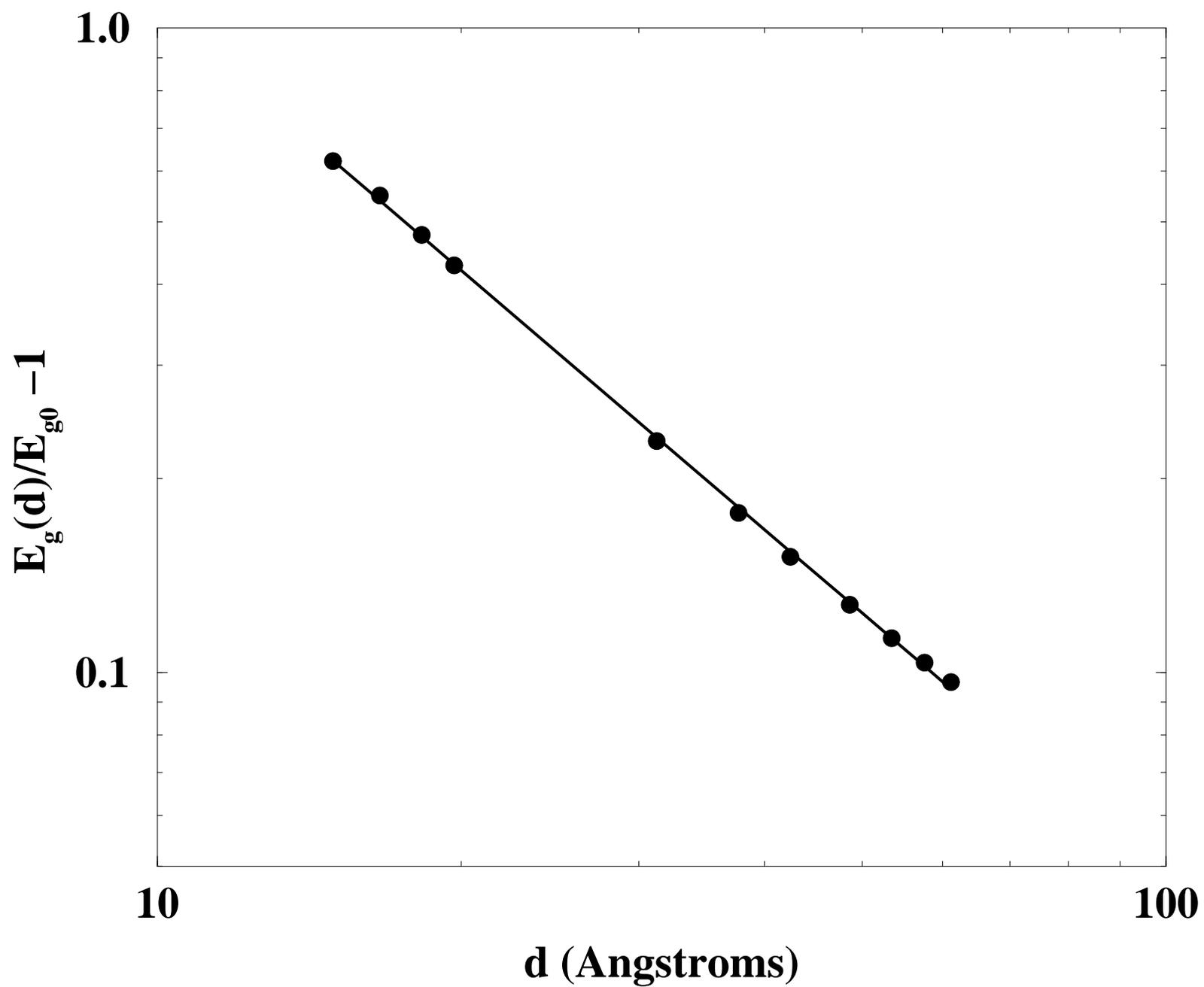
\end{document}